# Off-axis electron holography for the direct visualization of perpendicular shape anisotropy in nano-scale 3D magnetic random-access-memory devices

Trevor P. Almeida[1,2]*, Alvaro Palomino[3], Steven Lequeux[3], Victor Boureau[4], Olivier Fruchart[3], Ioan Lucian Prejbeanu[3], Bernard Dieny[3] and David Cooper[1]

[1]Univ. Grenoble Alpes, CEA, Leti, F-38000 Grenoble, France.
[2]SUPA, School of Physics and Astronomy, University of Glasgow, G12 8QQ, UK.
[3]Univ. Grenoble Alpes, CEA, CNRS, Grenoble INP, SPINTEC, 38000 Grenoble, France.
[4]Interdisciplinary Center for Electron Microscopy, EPFL, CH-1015 Lausanne, Switzerland.

Perpendicular shape anisotropy (PSA) and double magnetic tunnel junctions (DMTJ) offer practical solutions to downscale spin-transfer-torque Magnetic Random-Access Memory (STT-MRAM) beyond 20 nm technology nodes, whilst retaining their thermal stability and reducing critical currents applied. However, as these modern devices become smaller and three-dimensionally (3D) complex, our understanding of their functional magnetic behavior is often indirect, relying on magnetoresistance measurements and micromagnetic modelling. In this paper, we review recent work that was performed on these structures using a range of advanced electron microscopy techniques, focusing on aspects specific to the 3D and nanoscale nature of such elements. We present the methodology for the systematic transfer of individual SST-MRAM nano-pillars from large-scale arrays to image their magnetic configurations directly using off-axis electron holography. We show that improved phase sensitivity through stacking of electron holograms can be used to image subtle variations in DMTJs and the thermal stability of < 20 nm PSA-STT-MRAM nano-pillars during *in-situ* heating. The experimental practicalities, benefits and limits of using electron holography for analysis of MRAM devices are discussed, unlocking practical pathways for direct imaging of the functional magnetic performance of these systems with high spatial resolution and sensitivity.

*Corresponding author:
Tel: +44 (0) 141 330 4712
Email: trevor.almeida@glasgow.ac.uk

# 1. <u>Introduction</u>

Magnetic random-access memory (MRAM) is a type of non-volatile memory, based on the storage of individual bits of information by ferromagnetic cells with switchable polarity[1]. The ability to use an electric current to write magnetic states through spin-transfer torque (STT) has highlighted STT-MRAM devices as an industrially-relevant replacement for static RAM in on-chip cache memory[2,3]. This is owed to its low energy consumption[4], superior endurance[4,5], fast switching and ease of integration with complementary metal-oxide-semiconductor (CMOS) technology[6]. Conventional STT-MRAM cells use a magnetic tunnel junction (MTJ) comprising two ferromagnetic thin films separated by an insulating MgO tunnel barrier (1 – 1.5 nm thick). Perpendicular (p) STT-MRAM involves perpendicular magnetization, which is achieved through interfacial magnetic anisotropy, requiring films of thickness of few nanometers at most. The bottom layer acts as a reference layer, being magnetically pinned in a given direction perpendicular to its 2D plane with a synthetic antiferromagnet (SAF) layer, whilst the magnetic orientation of the top storage layer can be switched between respective up / down states by sufficiently strong current pulses[7,8]. The parallel and antiparallel alignment of the perpendicularly-magnetized layered film provides two states of significantly-different electrical resistance, and hence acts as a system of reading and writing '0' and '1' binary information[9].

To increase the areal bit density of modern p-STT-MRAM devices, there is a drive to reduce the in-plane lateral size or diameter of the corresponding MTJ[10]. However, below a characteristic magnetic length scale, the magnetic moment of the storage layer becomes excessively unstable to thermal fluctuations. One solution is to increase the out-of-plane aspect ratio of the storage layer by increasing its thickness to larger than its diameter, resulting in a dramatic increase in volume, while maintaining the perpendicular orientation of magnetization through perpendicular shape anisotropy (PSA) and not relying entirely on the interfacial effects[11,12]. This provides improved thermal stability of the three-dimensional (3D) STT-MRAM cells. Previous studies have shown that the PSA-STT-MRAM are indeed highly thermally stable, making them a practical solution to extend the scalability of STT-MRAM at sub-20 nm technology nodes[13,14].

Another solution to simultaneously improve the thermal stability while reducing the critical current required to switch the storage layer is using double MTJs (DMTJ). In a DMTJ stack, a secondary MgO tunnel barrier is sandwiched between the storage layer and an additional ferromagnetic thin film above, termed the polarizing layer. The improved thermal stability arises from the extra interfacial anisotropy contribution from the secondary MgO barrier. In specialized DMTJ systems, the polarizing layer can be switched to minimize or maximize the STT of the storage layer during the read and write modes, respectively. The functional efficiency of the DMTJs has been shown to be increased by a factor of 4 compared to conventional p-STT-MRAM cells[15] and requires much lower write currents[16]. In addition, post-deposition annealing can be used to further increase the tunnel magnetoresistance and perpendicular magnetic anisotropy of MTJs[17].

However, our knowledge of the local magnetic behavior of these various 3D p-STT-MRAM stacks is often indirect, relying on magnetoresistance measurements and micromagnetic modelling. In order to understand and optimize thermal stability, STT writability and magneto-resistive readability, it is necessary to image their magnetic configurations and the effect of temperature directly. Localized magnetic imaging can be ideally performed using a family of transmission electron microscopy (TEM) techniques collectively known as Lorentz microscopy, with modern aberration-corrected TEMs[18,19]. These techniques include Fresnel imaging[20,21], off-axis electron holography[22,23] and differential phase contrast (DPC) imaging[18,21]. In particular, off-axis electron holography allows imaging of magnetization within nanostructures of complex 3D geometry, with high spatial resolution approaching ~ 1 nm and sensitivity to induction field components transverse to the electron beam[24]. Combining electron holography with *in-situ* heating within the TEM has already allowed direct imaging of the thermal stability of nano-scale signal carriers and fields of magnetic minerals[25-27], meteorites[28], pre-patterned MTJ conducting pillars[29] and PSA-STT-MRAM nano-pillars[30,31]. However, there are several aspects that can make magnetic imaging of 3D nanoscale MRAM devices challenging, including: (1) sample preparation from the device substrates using dual-beam focused ion beam (FIB) secondary electron microscopes (SEM); (2) improving sensitivity and spatial resolution for samples with smaller interaction volumes; (3) reliable interpretation of two-dimensional (2D) representations of 3D magnetic components, and (4) amplification of the effects associated with *in-situ* TEM experiments at high magnifications, e.g. thermal drift.

In this paper, we present the methodology for the systematic transfer of individual SST-MRAM nano-pillars from large-scale arrays to image their localized magnetic configurations using off-axis electron holography. We demonstrate the ability to prepare single rows of SST-MRAM nano-pillars within the dual-beam FIB-SEM for TEM investigations with limited FIB exposure and irradiation. Through systematic acquisition of series of electron holograms, we reconstruct magnetic inductions maps with improved sensitivity to show the distribution of magnetization in ≤ 20 nm STT-MRAM nano-pillars. This allows for confirmation of their PSA and reveals subtle variations of magnetic orientation in thermally annealed multi-layered DMTJ nano-pillars. In addition, we experimentally demonstrate the influence of PSA on its thermal stability through *in-situ* heating within the TEM to 250 $^0$C. This study illustrates the ability to prepare a range of 3D STT-MRAM nano-pillars for electron holography with improved phase resolution and sensitivity, providing direct imaging of their PSA, subtle inter-layer interactions and thermal stability, that were previously inaccessible.

## 2. Materials and methods

A selection of three 2D arrays of PSA-STT-MRAM nano-pillars were fabricated through sequential e-beam lithography, reactive ion etching and ion beam etching[11]. The simplified stacks of PSA MTJs as they appear in this paper have the following compositions: $SiO_2$ substrate / buffer layer / SAF and/or FeCoB reference layer / ferromagnetic free layer(s) / capping layers / Ta mask. The detailed compositions of the ferromagnetic free layers that provide the PSA are displayed Figure 1. A Thermo Fisher Strata 400S FIBSEM was used for TEM sample preparation, with more details presented in section 3.1. High-angle annular dark field (HAADF) scanning TEM (STEM) imaging was performed using a Thermo Fisher Titan TEM at 200 kV, equipped with a probe aberration ($C_S$) corrector. Corresponding chemical analysis was provided by energy dispersive X-ray (EDX) spectroscopy using a Bruker Super-X system composed of four silicon-drift detectors. Off-axis electron holograms were acquired on a Gatan OneView 4K camera using a Thermo Fisher Titan TEM equipped and an electron biprism. In order to prevent the influence of the strong magnetic field of the conventional objective lens (in the order of 2 T), the electron holography was performed under field-free conditions in Lorentz mode. Image focusing was achieved with the Lorentz lens (objective mini-lens), and use of the $C_S$ image corrector allowed to maintain a spatial resolution approaching ~ 1 nm. Stacks of 8 electron holograms (each acquired for 4s) were aligned and averaged to improve the signal to noise ratio of the reconstructed phase images, in practice using the Holoview software[19]. Similarly, the Holoview software was used to perform π phase-shifting electron holography, to further improve the spatial resolution and signal-to-noise ratio of the phase. More details will be given in sections 3.2 and 3.3. *In-situ* heating up to 400°C was performed using a Gatan heating holder under field-free magnetic conditions to examine the effect of thermal annealing on stack 2, each time for 30 minutes. The additional thermomagnetic behavior of stack 3 was investigated up to 250°C, allowing 30 minutes to stabilize from thermal drift at every temperature interval. The heating was repeated, and the magnetic reversal was performed at each temperature interval to isolate the mean inner potential (MIP), and subtracted from the first heating to reconstruct the thermomagnetic behavior of the nano-pillars[25,26].

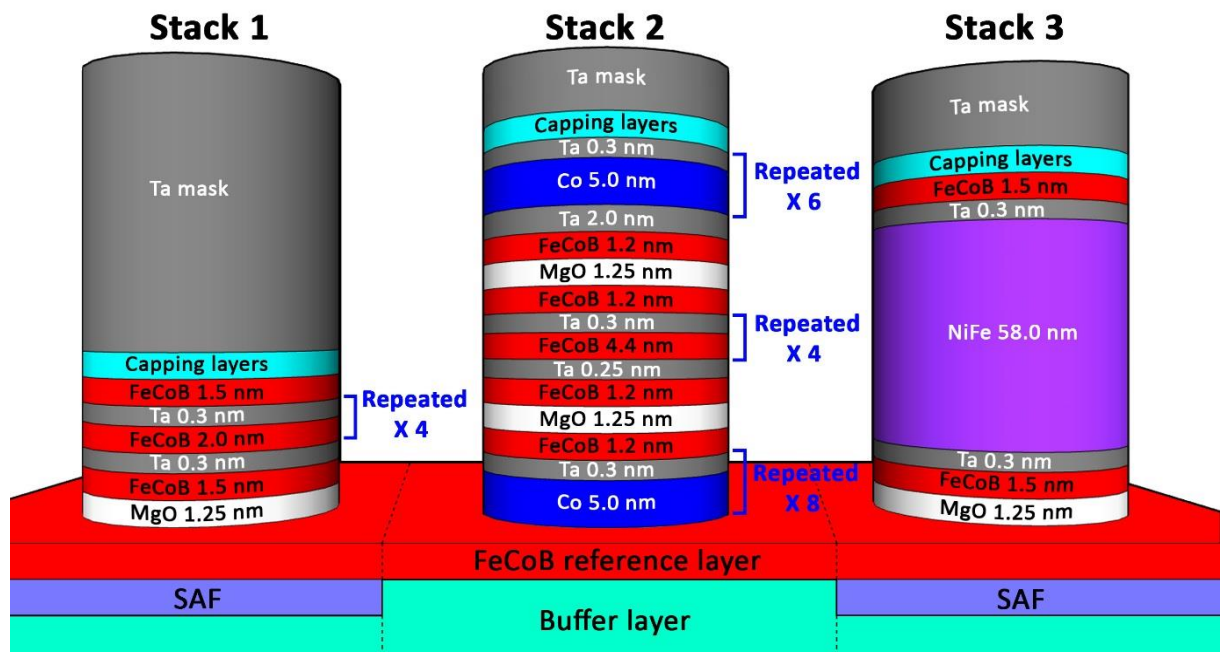

**Figure 1.** Schematic diagram of the three types of nano-pillar stacks, providing detailed compositions of the ferromagnetic free layers. Several consecutive layers of same dimension are repeated, as indicated.

### 3. Results

### 3.1 Sample preparation

Descriptions of sample preparation protocols are often neglected in reports of magnetic imaging using electron holography, although it is a critical step in the reliable reconstruction of their magnetic properties. Figure 2 presents the methodology for the systematic transfer and preparation of single rows of PSA-STT-MRAM nano-pillars from a large array of stack 1 for the purpose of their localized TEM characterization. This newly developed method allows reliable and repeatable TEM sample preparation of PSA-STT-MRAM nano-pillars with limited $Ga^+$ ion exposure, as $Ga^+$ ion implantation and irradiation can adversely affect their magnetic properties. The SEM image of Fig. 2a displays an exemplar array of PSA-STT-MRAM nano-pillars after the ion-beam etching stage, each of ~ 150 nm length and ~ 30nm diameter (Fig. 2a, inset), and with ~ 1 µm lateral separation (Fig. 2a, arrowed blue)[13]. To prevent against $Ga^+$ ion exposure, a ~ 1 µm thick layer of organic resin was spin-coated over the surface prior to transfer to the FIB-SEM. The sample was inserted in the FIB-SEM and tilted perpendicular to the FIB-beam direction (52°). A protective W strip (~ 20 µm long, ~ 3 µm wide, ~ 2 µm thick) was then deposited along the same axis as the rows of nano-pillars using a FIB current of 0.28 nA to protect the resin from $Ga^+$ ion irradiation and to provide mechanical stability. Large trenches were milled either side of the W mask (Fig. 1b) with progressively lower FIB currents (21 nA → 9.0 nA → 6.5 nA) to reduce the thickness of $Ga^+$-damaged surface layers. The FIB section (~ 18 µm long, ~ 2 µm wide, ~ 10 µm thick) was released and transferred to an Omniprobe® Cu lift-out grid, where the layers of protective W and organic resin layers are clearly observed (Fig. 2c). The front face of the section was tilted by 1° away from the FIB beam direction and milled using progressively lower FIB currents (0.92 nA → 0.46 nA → 0.28 nA) to further reduce the thickness of $Ga^+$-damaged surface layers until a row of nano-pillars were visible within the resin (Fig. 2d, inset). The section was tilted by 2° in the opposite direction (1° relative to its initial position) and back-milled using the same FIB currents until the lamella thickness was ~ 300 nm (Fig. 2e, inset). The Cu lift-out grid was then removed from the FIB-SEM and transferred to a plasma-cleaner for the purpose of etching the organic resin layer with an oxygen / argon plasma mixture (6 minutes). The plasma-etched lamella was re-inserted in the FIB-SEM, tilted to 52° to prevent exposure of the nano-pillars and a small notch (100 nm x 300 nm) was milled on the right-hand side of the remaining free-standing W protective mask (Fig. 2f, green arrow). The micro-manipulator was then positioned behind the left-hand side of the W mask, before being driven forward, bending the W mask ~ 5 µm (Fig. 2f, white arrow direction) whilst only monitoring the motion of the W mask with the e-beam. The micro-manipulator was then driven rightwards, further bending the W mask away from the lamella or until it breaks at the FIB notch (Fig. 2g), leaving a single row of nano-pillars for the purpose of TEM analysis (Fig. 2h). The HAADF image and EDX chemical maps of Fig. 2i display an exemplar nano-pillar (stack 1) comprising a magnetic Co / Fe section (bottom) with a diameter as small as ~ 10 nm. The chemical analysis also reveals the thin Ru capping layer, hard Ta mask (~ 15 nm diameter, ~ 70 nm length) and external O layer, attributed to both residual organic resin and surface oxidation.

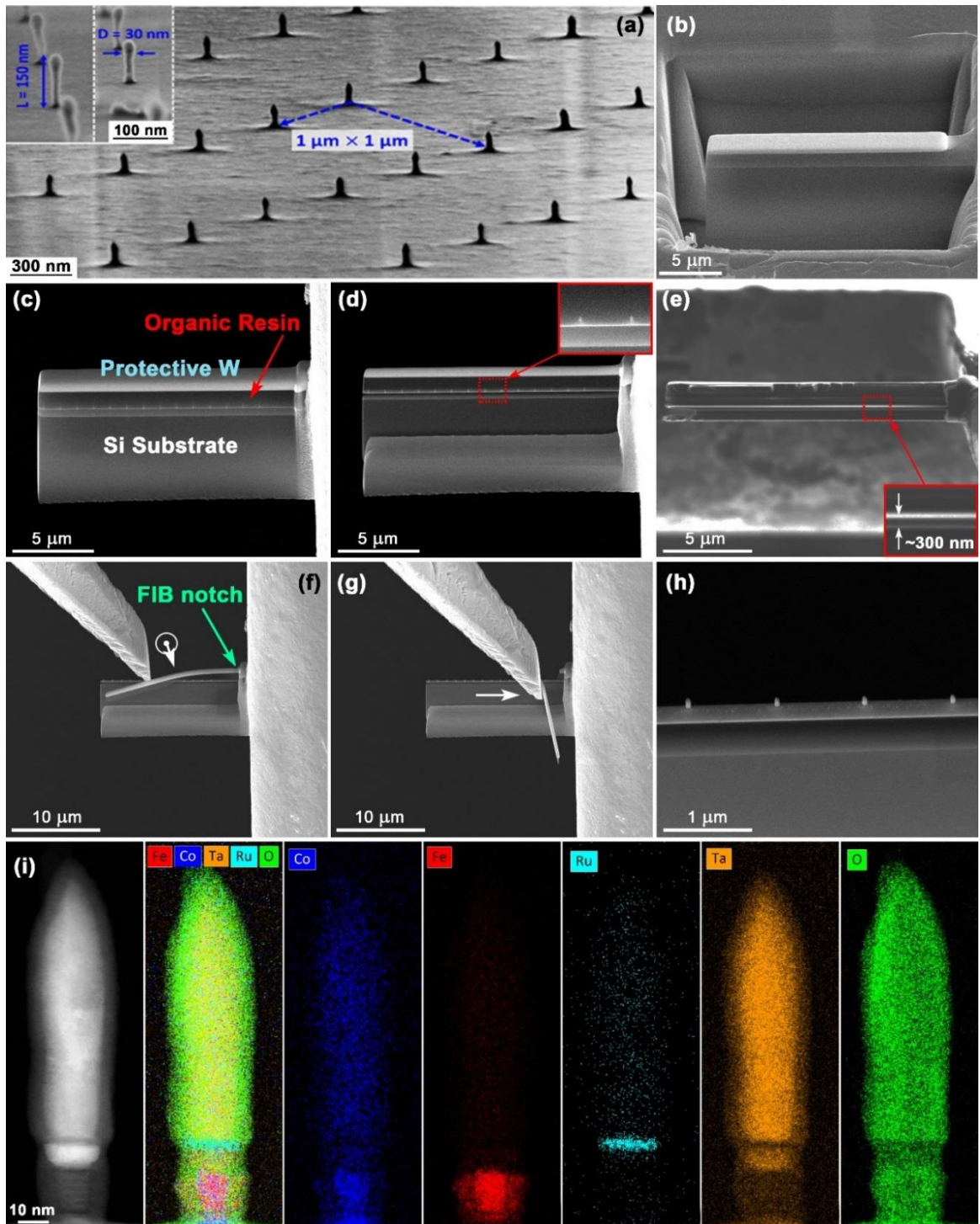

**Figure 2.** (a) SEM image of an array of PSA-STT-MRAM nano-pillars (~ 150 nm length and ~ 30nm diameter) fabricated through sequential ion beam etching. Adapted from Ref[13]. (b-d) SEM images of the (b) large FIB-milled trenches either side of the FIB-deposited protective W mask; (c) lamella transferred to a Cu lift-out grid showing the protective W mask, organic resin layer and Si substrate; and (d) front-milling of the lamella with sequentially smaller FIB currents to reveal the desired row of nano-pillars (inset). (e) FIB image of the back-milled lamella with a thickness of ~ 300 nm (inset). (f-h) SEM images showing the FIB notch position and motion of the micro-manipulator (arrowed, white) to (f) bend; and (g) break the W mask; to (h) reveal a single row of nano-pillars. (i) HAADF STEM image and EDX chemical maps of an exemplar nano-pillar (stack 1) showing its morphology and Co, Fe, Ru, Ta and O content.

## 3.2 Magnetic imaging using electron holography

Off-axis electron holography was used to image the magnetic configuration within the ~ 20 nm diameter PSA-STT-MRAM nano-pillars. Figure 3 presents a schematic of the basic experimental setup for the electron holography of the nano-pillar samples prepared in section 3.1. A voltage is applied to the electron biprism (Fig. 3a), which allows two components of a coherent electron beam to interfere: (1) the objective wave that passes through the sample; and the (2) reference wave passing near the sample, through a region free of electromagnetic potential. The overlap region of the two interfering waves is typically between 100 nm and 2 µm wide, and is proportional to the applied biprism voltage, but inversely proportional to the interference fringe spacing and fringe contrast. For this study of small PSA-STT-MRAM nano-pillars, voltages in the range 150 V to 220 V were applied to the biprism and resulted electron holograms with interference fringe spacings from ~ 2.3 nm down to ~ 1.6 nm, respectively. The intensity and position of the interference fringes were used to reconstruct both amplitude and phase information of the nano-pillars, respectively, by using a processing routine based on fast Fourier transforms (FFTs) (Fig. 3b). The magnetic properties of the nano-pillars are obtained from the magnetic contribution to the reconstructed phase image. The spatial resolution of the phase image is directly related to the interference fringe spacing while the phase sensitivity is related to the fringe contrast, acquisition time and electron beam intensity. The high spatial coherence required for electron holography is achieved by significantly elongating the electron beam in the direction perpendicular to the electron biprism axis, with the large directional spread resulting in higher coherence of the electrons either side of the biprism. Consequently, the elliptical illumination reduces the total amount of electrons available to create the interference fringes of the overlap region. Longer acquisition times on the order of several tens of seconds increases the signal-to-noise ratio in the recorded electron hologram and hence improves the phase sensitivity. However, the microscope is more susceptible to instabilities and sample drift during these long acquisitions, particularly at the high magnifications of 70,000 times used for the small PSA-STT-MRAM nano-pillars and during *in-situ* heating experiments. As the electron beam intensity is limited, the alternative method to improve the signal-to-noise ratio of the phase image is to acquire a series of holograms and add them together, through which phase sensitivities of $2\pi/1000$ rad have been reached[31-33]. The fringe contrast in the electron holograms and spatial resolution of phase maps can be further improved by π phase-shifting holography, where the electron beam was tilted in order to shift the holograms by half the interference fringe spacing between successive hologram acquisitions, such that the center band could be removed from the Fourier reconstruction by subtraction of the two holograms[34]. These improvements will be demonstrated in section 3.3. The total phase shift, $\phi(x,y)$, of the electron wave recorded perpendicular to the incident electron beam direction $z$ has two components sensitive to (1) the electrostatic potential, $\phi_e$, and; (2) the out-of-plane component of the magnetic vector potential, $\phi_m$, in the specimen, as summarized by:

$$\phi(x,y) = \phi_e(x,y) + \phi_m(x,y) = C_E \int V(x,y,z)\, dz - \left(\frac{2\pi e}{h}\right) \int A_z(x,y,z)\, dz$$

where $C_E$ is an interaction constant at the chosen TEM accelerating voltage, $V$ is the electrostatic potential, $e$ is the elementary charge of an electron and $A_z$ is the out-of-plane magnetic vector potential. The $\phi_m$ can be isolated through acquiring two separate electron hologram sets before and after reversing the magnetic configuration of the sample (thus the sign of $A_z$ and $\phi_m$), subtracting one $\phi$ from the other to eliminate the $\phi_e$ contribution, and dividing by two. Reversing the sign of $\phi_m$ is commonly achieved by two methods: 1) physically flipping the sample by 180° either inside or outside the TEM; or 2) tilting the sample to large opposing tilt angles whilst positioned between the TEM pole-piece and applying saturating magnetic fields using the TEM objective lens, with the assumption that the sample relaxes into perfectly opposite magnetic states. Both these methods are implemented in this paper. A less common method is to use a separate controllable magnetic source within the TEM or sample holder to apply in-plane fields to reverse the magnetization component of the sample.

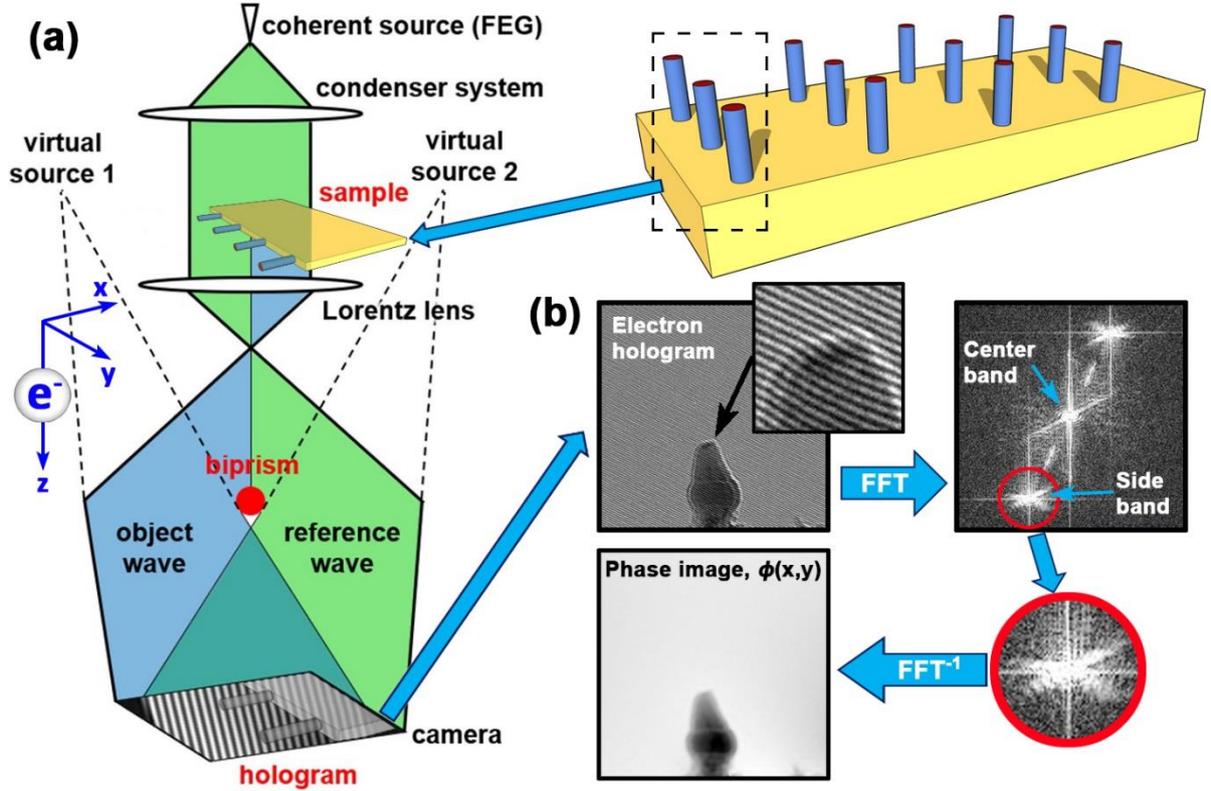

**Figure 3.** (a) Schematic of the experimental setup for off-axis electron holography. (b) Procedure for using an electron hologram to reconstruct a total phase image, $\phi(x,y)$, of a small nano-pillar.

Figure 4 illustrates the process for reconstructing the $\phi_m$ and associated magnetic induction maps of an exemplar nano-pillar from stack 2. The amplitude image of Fig. 4a displays the general morphology of the nano-pillar with a diameter of ~ 40 nm at its largest. Fig. 4b and Fig. 4c demonstrate the variation in the $\phi(x,y)$ maps of the nano-pillar in both orientations after tilting the sample and applying a saturating field in opposite directions using the objective lens. The asymmetric part in each phase map is attributed to the $\phi_m$ contribution from the stray magnetic field, appearing as contributions with opposite intensities on either side of the nano-pillar. The $\phi_m$ is isolated by the subtracting the two orientations (Fig. 4d) and the cosine of $\phi_m$ (amplified by a factor of 100) allows the visualization of magnetic equipotential lines evidencing the magnetic contours (Fig. 4e). It clearly shows the magnetization of nano-pillar orientated along its long axis, as well as a stray magnetic field emanating from its tip. The map of the in-plane magnetic induction $B_\perp$ shown in Fig. 4f is calculated from the phase gradient of $\phi_m$ using the following equation:

$$\boldsymbol{B}(x,y,z) = \nabla \times \boldsymbol{A}(x,y,z)$$

where $\nabla$ is the nabla operator and $\times$ is the cross product. The directional information of the in-plane induction can then be included and is indicated by the color wheel (Fig. 4f, inset).

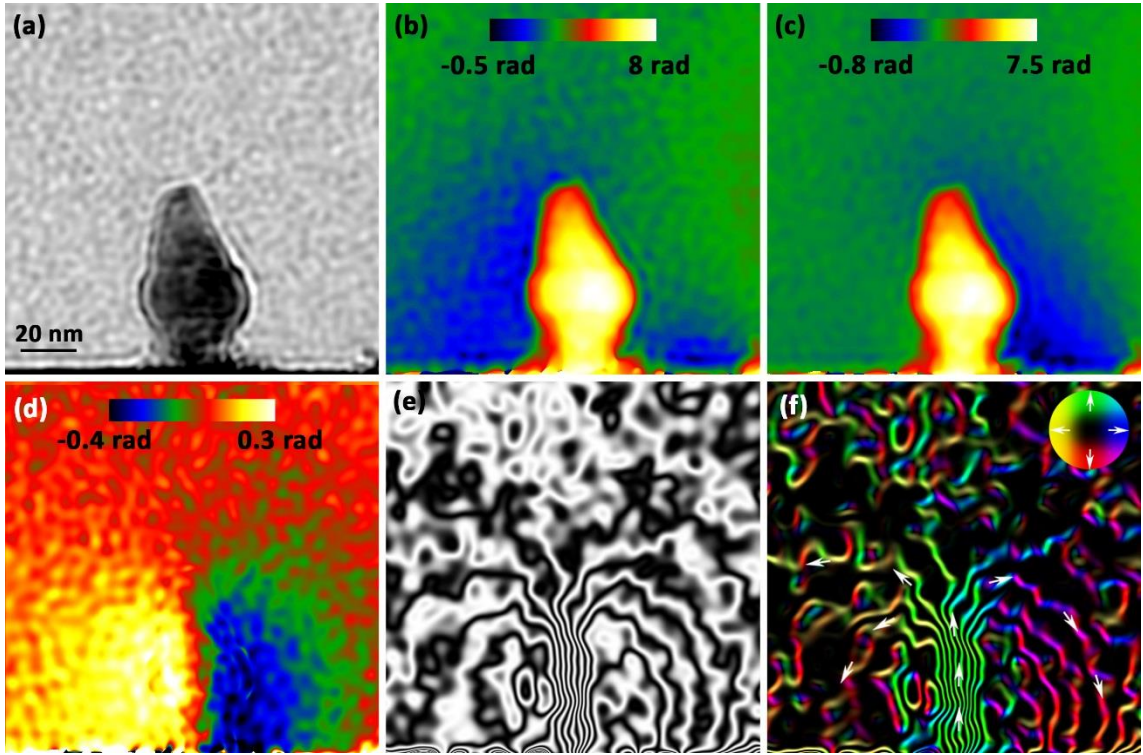

**Figure 4.** (a-c) Electron holography of the exemplar nano-pillar (stack 2) and its reconstructed (a) amplitude; and (b,c) phase images after applying a saturating field in opposite directions by tilting to ±40°. (d) The isolated magnetic contribution to the phase, $\phi_m$. (e) Amplified cosine of (d) by 100 times to create a map of equipotential lines. (f) Projected magnetic induction map calculated from the phase gradient of (d), with direction of magnetization indicated by the color wheel (inset).

### 3.3 Magnetic imaging of the PSA-STT-MRAM nano-pillars

Figure 5 illustrates how electron holography can be used to image subtle magnetic variations of a multi-layered nano-pillar, by improving the $\phi_m$ signal through image stacking. The HAADF STEM image displays its morphology (~ 60 nm height, ~ 35 nm largest diameter) and layered growth structure of stack 2 (Fig. 5a). The associated EDX chemical maps (Fig. 5b) shows clearly that the nanopillar comprises a multi-layered composition of Co (40 nm) / MgO (2 nm) / FeCoB (20 nm) / MgO (2 nm) / Co (20 nm), as expected from the schematic of stack 2 displayed in Figure 1. Similar to Figure 2i, the external O layer is attributed to both residual organic resin and surface oxidation. The reconstructed $\phi_m$ of Fig. 5c reveals that magnetization is oriented along the long axis of the nano-pillar and the line profile taken across its diameter indicates the level of $\phi_m$ signal-to-noise in single holograms (Fig. 5e, blue line). Fig. 5d demonstrates the significant improvement in $\phi_m$ signal from the small nano-pillar achieved by aligning and averaging an 8-image stack of holograms, which is reinforced by the analogous line profile (Fig. 5e, red line). The corresponding contoured magnetic induction map (Fig. 5f) is created through combining amplification of the cosine (×100) and the phase gradient to indicate field direction (colour wheel, inset). It shows clearly that magnetization flows through the multi-layered nanopillar in a direction parallel to its long axis (indicated by white arrows), confirming its out-of-plane PSA, and gives rise to a stray dipolar magnetic field. Figs 5g-i illustrate the effect of thermal annealing on the magnetic configuration of the native nano-pillar (Fig. 5g), all acquired at room temperature, post-annealing. After initial annealing to 400°C, the magnetic contours flow in a transverse direction across the central FeCoB layer and deviate away from the long axis in the top Co layer (Fig. 5h). This change in magnetic orientation becomes more pronounced with additional *in-situ* annealing to 400°C (Fig. 5i) and the understanding of this effect will be described in the next section.

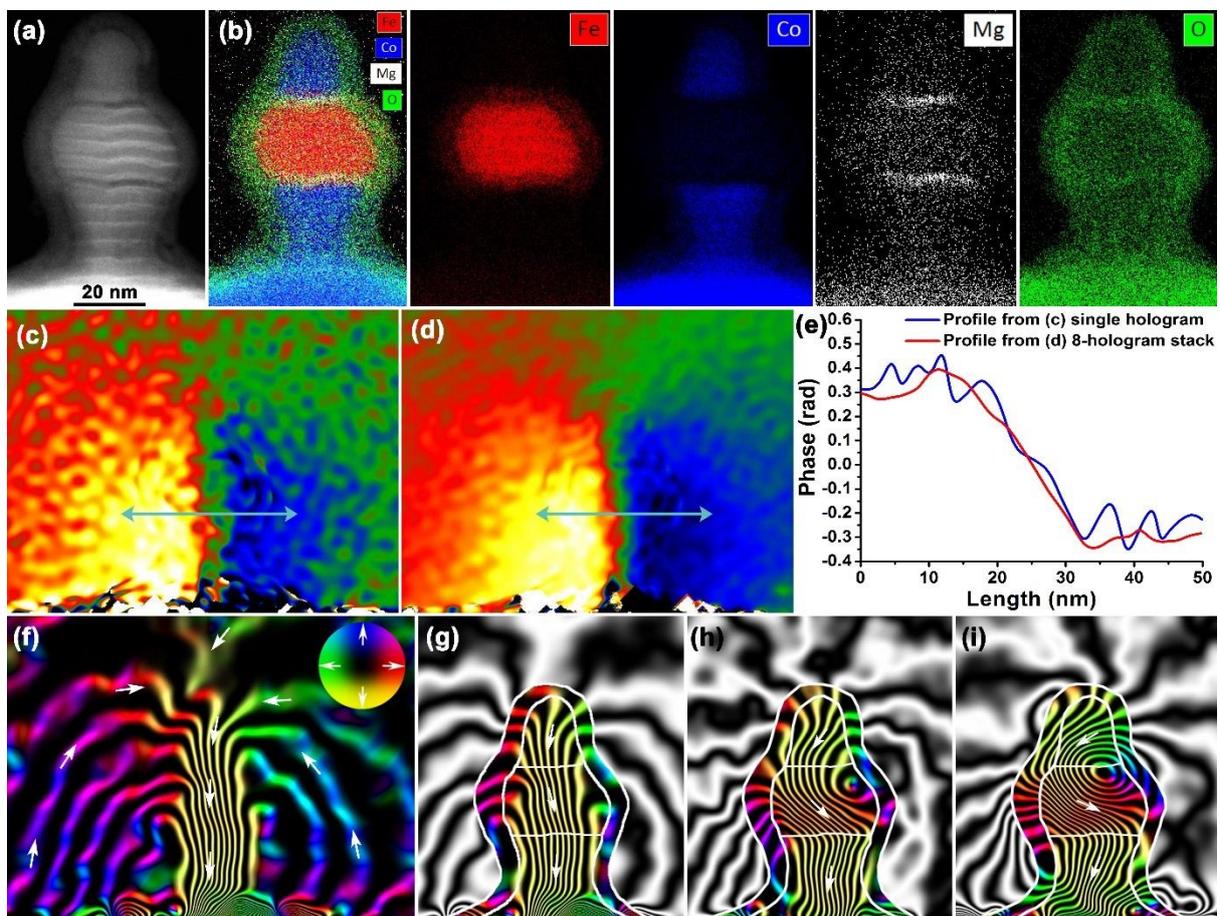

**Figure 5.** (a) HAADF STEM image of a multi-layered nano-pillar from stack 2 (~ 60 nm height, ~ 35 nm largest diameter); and (b) EDX chemical maps showing the elemental distribution of Fe, Co, Mg and O. (c) Reconstructed $\phi_m$ contribution of the nano-pillar from a single electron hologram. (d) $\phi_m$ signal by averaging an 8-image electron hologram stack. (e) Graph of the line profiles taken across the diameter of the nano-pillar in (c, blue) and (d, red), demonstrating a significant improvement in $\phi_m$ signal-to-noise ratio. (f) Associated magnetic induction map showing contours flowing along the major axis of the multi-layered nanopillar and a dipolar stray magnetic field. The contour spacing is 0.063 rad (cosine of 100 times the $\phi_m$) and the field direction is indicated by the color wheel (arrowed, inset). (g-h) Magnetic induction maps (g) before; and after (h) first; (i) secondary *in-situ* annealing to 400 °C for 30 minutes.

Figure 6 presents the thermomagnetic behavior of a FeCoB / NiFe nano-pillar (stack 3) during *in-situ* heating to 250 $^0$C. The HAADF STEM image (Fig. 6a) displays the general morphology of the nano-pillar with a Ta mask (top) and lower shaft. This is confirmed by EDX chemical mapping of Fig. 6b, revealing that the NiFe section of the nano-pillar is ~ 60 nm high with a diameter of ≤ 20 nm, and is separated from the hard Ta mask by the ~ 3 nm Ru capping layer. The initial combined EDX / magnetic induction map of Fig. 6c is created from a 8-hologram series and displays the nano-pillar at 20 $^0$C. The magnetization is clearly pointing from the bottom to the top of the NiFe section and confirms its PSA. The associated magnetic induction maps of Fig. 6c display the nano-pillar during heating from 20 $^0$C to 250 $^0$C. It is evident that the NiFe section retains its upward direction of magnetization at all temperatures up to 250 $^0$C.

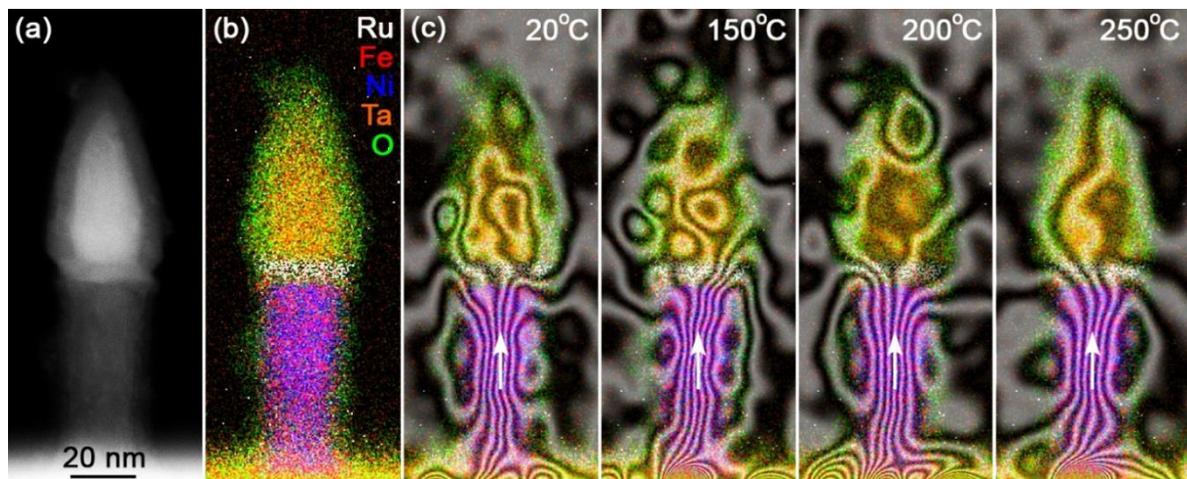

**Figure 6.** (a) HAADF STEM image of a nano-pillar from stack 3; and (b) associated EDX chemical map showing the elemental distribution of Ru, Fe, Ni, Ta and O. (c) Combined EDX / contour maps showing magnetic induction contours flowing along the major axis of the ~ 20 nm diameter NiFe section (pink), at 20 °C and during *in-situ* heating to 150 °C, 200 °C and 250 °C. The contour spacing is 0.042 rad (cosine of 150 times the $\phi_m$) and the field direction indicated by arrows.

Figure 7 presents the morphology, chemical analysis and electron holography of a larger nano-pillar (stack 3), with the HAADF image of Fig. 7a showing its diameter of ~ 70 nm. The associated EDX analysis (Fig. 7b) displays the magnetic NiFe section to exhibit a height of ~ 60 nm, which is again separated from the hard Ta mask with a thin Ru layer, as well as an external O layer attributed to surface oxidation and residual organic resin. Fig. 7c shows a single electron hologram of the thicker nano-pillar, where the interference fringes are difficult to resolve in the Ta mask, which will influence the phase reconstruction from the FFT (Fig. 7c, inset). In order to improve the contrast of the interference fringes, the π phase-shifting electron holography method was used to obtain the hologram presented in Fig. 7d. In this case, the interference fringes are phase-shifted by π in two consecutive electron holograms, and their difference shown in Fig. 7d demonstrates the removal of the contribution of the bright-field image to the electron hologram and hence the removal of the center band in the FFT. This allows a reduction in voltage applied to the biprism to increase the interference fringe contrast, but without losing spatial resolution thanks to the ability to use a larger numerical aperture on the sideband (acting as a low-pass filter) during the FFT reconstruction process (Fig. 7d, inset)[34]. The magnetic contour map (Fig. 7e) was reconstructed from two π phase-shifted 16-electron hologram series, with the magnetization of the nano-pillar in each stack reversed by physically flipping the sample. After application of the π phase-shifting method on 16-electron hologram series, series of 8 phase images are obtained and then aligned to provide averaged phase images with improved signal and spatial resolution. The difference between the two average phase images did present artefacts in the Ta mask and hence only the magnetic induction present in the NiFe section is shown in Fig. 7e. Nevertheless, the magnetic signal is easily resolved and is consistent with a vortex-state, where the core of the vortex is oriented along the major axis of the nano-pillar (Fig. 7e).

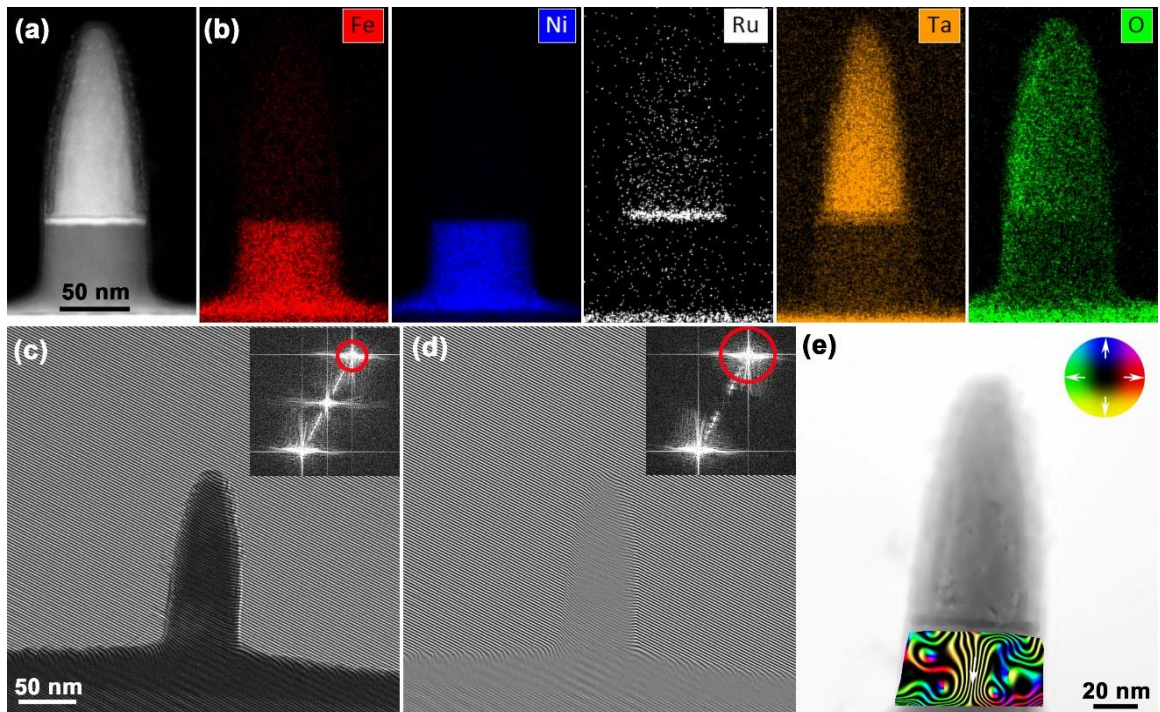

**Figure 7.** (a) HAADF STEM image of a large PSA-STT-MRAM nano-pillar (~ 200 nm height, ~ 70 nm largest diameter); and (b) EDX chemical maps showing the elemental distribution of Fe, Ni, Ru, Ta and O. (c) Electron hologram of the large nano-pillar and associated FFT (inset) with aperture used for phase reconstruction (red circle). (d) Difference between two consecutive electron holograms phase-shifted by π, with associated FFT (inset) and larger aperture used for phase reconstruction (red circle). (e) Magnetic induction map showing contours flowing along the major axis of the NiFe section. The contour spacing is 0.126 rad (cosine × 50 times) and the field direction is indicated by the color wheel (arrowed, inset).

## 4. Discussion

In this paper we have shown that it is becoming routine to prepare and transfer single rows of < 10 nm diameter p-STT-MRAM nano-pillars from large arrays with minimal FIB beam damage for localized TEM analysis. However, their small size and volume makes access to their local magnetic properties using TEM techniques challenging. Here, off-axis electron holography is demonstrated to be a very effective method to visualise the magnetic configuration of nano-pillars with diameters as small as ~ 20 nm. This methodology is enhanced through averaging and π phase-shifting series of electron holograms, improving both the sensitivty and spatial resolution of the reconstructed phase images. By averaging only 8 electron holograms, the signal-to-noise ratio observed for the stray field is experimentally doubled, which is key for the magnetic mapping of these small devices. If required, a larger culmulative number of holograms can be summed to compensate for the low beam intensities used for off-axis electron holography. Isolation of the magnetic contribution to the phase, $\phi_m$, has confirmed the out-of-plane PSA of the nano-pillars which has only previously been measured indirectly by magnetoresistance measurements or simulated through micromagnetic modelling. Further, combining electron holography with *in-situ* heating within the TEM has provided a direct route to image subtle variations in their magnetic configurations and their thermal stability. These aspects are less explicit in indirect bulk magnetoresistance measurements, which cannot distinguish physics such as a decrease of the tunneling magneto-resistance effect, a decrease of the magnitude of magnetization or a tilt of magnetization direction. This is demonstrated in Figs 5 and 6 where the *in-situ* heating of the nano-pillars exhibited significantly different magnetic behaviors. The thermal annealing of the multi-layered nano-pillar to 400°C (Fig. 5) has clearly influenced the interaction between the magnetic layers. The nano-pillar initially exhibited PSA with magnetization oriented along its long axis through all three layers. However, the higher temperature of 400°C is considered to have chemically altered the multi-layered structure through diffusion and intermixing, which has degraded both the magnetic properties of individual layers and their exchange interaction across their interfaces. For example, the coercivity, unidirectional anisotropy coefficient and saturation magnetization of CoFeB has been shown to decrease when annealing to > 300 °C[35]. Annealing to > 400 °C has also been shown to cause intermixing within Co-alloy layered films, resulting in a 40% reduction of its exchange bias field due to a decreased mean exchange stiffness across their interfaces and mean ferromagnetic interfacial moment[36,37]. Further, Ta has been shown to be susceptible to interdiffusion at high temperatures which can further dilute the saturation magnetization of CoFeB / Co layers[17]. All these factors combined with the middle CoFeB layer exhibiting a larger diameter than thickness, which provides a degree of shape anisotropy along a transverse axis of the nano-pillar, is thought to cause the degradation of the initial PSA across the multi-layered nano-pillar. In contrast, the heating of the FeCoB / NiFe nano-pillar (Fig. 6) reveals that the dipolar magnetic orientation along the major axis of the NiFe section (bottom to top) is retained at all temperature intervals up to at least 250 $^0$C. This thermal stability is significantly larger compared to standard p-STT-MRAM stacks based on ultrathin films[38-41], supporting the use of PSA-STT-MRAM in a variety of applications that require reliable performance over a range of operating temperatures. However, an increase in the diameter of NiFe section (Fig. 7) to values analogous to its layer thickness (~ 70nm) changes the favored magnetic configuration. Fig. 7e shows that flux loops on either side of central magnetic contours are contained within the NiFe section; indicative of a vortex state observed edge-on[42-44]. Hence, it is interpreted that the vortex core aligns with the long axis of the nano-pillar, with the reference FeCoB layer providing a degree of perpendicular anisotropy at the interface before wrapping around the vortex core in the NiFe section. Whilst the larger diameter of 70 nm may not be ideal for increasing the areal bit density of SST-MRAM devices, binary switching of vortex cells using alternating currents have been explored in vortex random-access memory (VRAM)[45]. Vortex states are also of great interest in magnetic sensor devices[46], oscillators[47] and biomedical devices[48]. Due to the 3D nature of the vortex state and interfacial coupling with FeCoB layer, it necessary to be aware that tilting the sample and applying the magnetic field of the TEM objective lens may not consistently reverse its more complex magnetic configuration compared to the previous nano-pillars (Figs 5 and 6). On the other hand, whilst physically flipping the sample in Fig. 7 will reverse the projection of $\phi_m$, care must be taken due to artefacts introduced from thick Ta mask, alignment of the opposite $\phi_m$ images and the distortions in protector lens system[49]. A lesser caveat of performing electron holography or *in-situ* heating of these nano-pillars is the presence of the organic resin residual from sample preparation. It is advised to perform an initial heating to induce any chemical changes to both the sample and resin layer to prevent further changes during subsequent heating of the experiment. As the residual organic resin is non-magnetic it did not contribute to the $\phi_m$ images, but excess amounts may produce electrostatic charging and introduce artefacts. Wet-chemical etching of FIB lift-outs with HF acid has been shown to produce high-quality Si/SiGe nano-pillars with clean oxide free surfaces[50]. However, the use of HF acid is known to preferentially react with Ni to form an insoluble $NiF_2$ layer[51] and hence would degrade the structure and magnetic properties of the NiFe section.

In conclusion, we have provided a direct route for the localized magnetic studies of a range of 3D p-STT-MRAM nano-pillars. We have shown that electron holography, through acquisition of hologram series or π phase-shifting, can be used to image directly subtle variations or thermal stability of < 20 nm PSA-STT-MRAM or DMTJ nano-pillars during *in-situ* heating. As smaller and more complex p-STT-MRAM devices are fabricated, this study unlocks practical pathways for direct imaging of the functional magnetic performance of these 3D systems with high spatial resolution and sensitivity.

**Funding**

CEA-Leti is a Carnot Institute. This work was supported by the French ANR via Carnot funding and the European Research Council (ERC MAGICAL, 664209).